\documentclass[%
 reprint,
 amsmath,amssymb,
 aps,
prb,
longbibliography,
superscriptaddress,
]{revtex4-2}

\usepackage{graphicx}
\usepackage{amsfonts,amsmath,amssymb}
\usepackage{relsize}
\usepackage{physics}
\usepackage{url}
\usepackage[normalem]{ulem}
\usepackage{xcolor}
\usepackage{comment}

\newcommand{\nano}{Centro S3, CNR–Istituto Nanoscienze, I-41125 Modena, Italy}
\newcommand{\unimore}{Dipartimento di Scienze Fisiche, Informatiche e Matematiche, Università di Modena e Reggio Emilia, I-41125 Modena, Italy}

\begin{document}

\title{Ab initio study of saddle-point excitons in monolayer SnS$_2$}

\author{Vinicius Alves Bastos}
\email[corresponding author: ]{vinicius.alvesbastos@unimore.it}
\affiliation{\unimore}
\author{Fulvio Paleari}
\email[corresponding author: ]{fulvio.paleari@nano.cnr.it}
\affiliation{\nano}
\author{Eleonora Luppi}
\affiliation{Laboratoire de Chimie Théorique, Sorbonne Université, CNRS, Paris F-75005, France}
\author{Marco Gibertini}
\affiliation{\unimore}
\affiliation{\nano}
\author{Alice Ruini}
\affiliation{\unimore}
\affiliation{\nano}

\begin{abstract}

Monolayer SnS$_2$ has emerged as a promising visible-light photocatalyst for photoelectrochemical applications, owing to its strong optical absorption in the visible range and excellent chemical stability. 
Despite its reduced dimensionality -- where excitonic effects are expected to be pronounced -- comprehensive theoretical investigations of bound excitons in this material remain scarce. Notably, unlike most two-dimensional hexagonal crystals, monolayer SnS$_2$ exhibits its lowest single-particle transition at the M point of the Brillouin zone (BZ). 
Here, the electronic valence bands form a saddle point while conduction states display a minimum with pronounced anisotropy, creating a distinctive band topology whose impact on optical excitations has not yet been systematically explored. 
In this work, we present a first-principles study of bound excitons in monolayer SnS$_2$ based on state-of-the-art many-body perturbation theory, employing the GW approximation and the Bethe–Salpeter equation (BSE). 
We analyze how band symmetry and anisotropy shape the excitonic wavefunctions and transition dipole moments. 
By resolving the exciton dipoles in momentum space for different linear light polarizations, we demonstrate that linearly polarized light lifts the $C_3$ rotational symmetry relating the three inequivalent M points, giving rise to three linearly independent excitonic states. 
This polarization-selective coupling, previously identified for saddle points in graphene, is achieved in SnS$_2$ for bound excitons and provides a potential route toward state encoding schemes in valleytronics applications.
\end{abstract}

\maketitle

\section*{Introduction}  

Layered materials are crucial building blocks for next-generation technologies such as solar cells, fuel cells, light-emitting diodes, batteries, sensors, memories, and quantum circuits \cite{mas+11nsc,xu+13cr,gupt+15pms,scha+16natrm,manz+17natrm,liu+22mlab}.
Their structure of stacked ultrathin atomic layers weakly bonded along the out-of-plane direction enables device miniaturization and flexibility, through few-layer exfoliation \cite{sun+12ac,sark+23as} or controlled growth \cite{sutt-sutt18acsanm,sutt+20cm}, as well as the design of new functional materials by stacking of different layered components \cite{hu+13mrb,cao+18acsaem,zrib+19npj2dma,sutt+23nano}. 
Additionally, reduced dimensionality often leads to remarkable physical features \cite{huan+14nano,scha+16natrm,wu-zheng16nl,wang-qian17nl,manz+17natrm,liu+22mlab}, including large in-plane charge mobility \cite{huan+14nano,kuma+21jpcc}, strongly bound excitons due to reduced dielectric screening and electron-hole confinement \cite{feng+15prb,manz+17natrm}, and unconventional topological behavior \cite{scha+16natrm,manz+17natrm,liu+22mlab}.
Among 2D materials, monolayer SnS$_2$ has attracted great attention as a promising visible-light photocatalyst for hydrogen production \cite{sun+12ac,zhon+12jpcc,hu+13mrb,cao+18acsaem}.
For example, Sun {\it et al.} reported remarkable incident photo-to-current efficiencies (IPCE), up to $38.7\%$ in the violet range (420 nm), for a photoelectrochemical cell (PEC) device based on SnS$_2$ monolayers \cite{sun+12ac}. 
SnS$_2$ exhibits a hexagonal Bravais lattice and its most stable allotrope is the T phase, characterized by an octahedral coordination of the Sn atoms, and lying 30 $k_B T$ (ambient conditions) below the metastable H phase~\cite{sun+12ac,zhua+13prb}.
The absence of unpaired electrons makes SnS$_2$ pristine crystals  non-magnetic, as experimentally confirmed in Ref.~\cite{li+17natc}. 
In fact, in the same work \cite{li+17natc}, they show that ferromagnetic ordering can be achieved through Fe doping, expanding its possible range of applications in nanodevices.
SnS$_2$ also combines several features that make it quite appealing for technological applications: low-cost synthesis, chemical stability in aqueous solution, non-toxicity,  
earth-abundancy, an experimental optical gap of 2.1 to 2.2 eV with a high absorption coefficient, and exfoliability down to a single layer \cite{sun+12ac,huan+14nano,eads+17natc}.
Despite these favourable characteristics, only a few theoretical studies have addressed the optical properties of monolayer SnS$_2$ \cite{zhua+13prb,baca+16prb,gonz-oley16prb}, and excitonic effects have been explicitly accounted for in only one \cite{zhua+13prb}.
Therefore, a comprehensive and accurate description of its bound excitons is still missing, even though such information is crucial for interpreting its optical response and assessing its potential for optoelectronics.
In this regard, in recent years, a few theoretical works have predicted promising new features in 2D hexagonal crystals, driven by optical excitations near the M-point saddle \cite{avet+22pra,shar+25acsnano,nual+25arxiv}.
Although the M saddle is a common feature in hexagonal lattices, the fact that it corresponds to the direct band gap (E$_{\text{dir}}$) in monolayer SnS$_2$ \cite{zhua+13prb,baca+16prb,gonz-oley16prb} is not common, and, as we show here, leads to the formation of strongly anisotropic bound excitons near this point with special symmetry properties.

In this work, we provide a detailed analysis of bound excitons in T-phase monolayer SnS$_2$ using many-body perturbation theory, consisting of the GW approximation and BSE.
We identify a richer spectrum of bound excitons than previously reported \cite{zhua+13prb}, including strongly bound dark excitons and several bright states in the 2.3–3.0 eV range.
Remarkably, the transition involves states around the M point in the BZ that correspond to a saddle point for the valence band and a minimum for conduction states, creating a distinctive situation rarely explored in the literature.
Furthermore, by analysing the momentum-resolved exciton-light coupling matrix elements under different light polarization directions, we demonstrate that linearly polarized light can selectively excite the three inequivalent M points, splitting each bright exciton into three linearly independent excitonic states, with potential for state representation in quantum electronics. 
While an analogous effect has been recently predicted in graphene \cite{shar+25acsnano} to enable ``saddletronics'', the robust nature of M-point excitons in SnS$_2$ opens interesting perspectives in the more established field of valleytronics \cite{seyl+26arxiv}.

\section{Methodology}
\label{methodology}

The optical excitations of monolayer SnS$_2$ are described within the framework of many-body perturbation theory, which was applied starting from a ground-state electronic structure, as obtained by means of Density-Functional Theory (DFT). 
The quasiparticle corrections to the DFT electronic band structure are computed within the one-shot GW approximation, followed by the solution of the BSE for the correlated electron-hole Hamiltonian.
The interacting quasiparticle (QP) energies $E^{\mathrm{QP}}_{n\mathbf{k}}$ are obtained from the perturbative solution of the Dyson equation \cite{sang+19jpcm,golz+19fic},
\begin{equation}
    E^{\mathrm{QP}}_{n\mathbf{k}}
    =
    \varepsilon^{\mathrm{DFT}}_{n\mathbf{k}}
    +
    Z_{n\mathbf{k}}
    \,
    \langle n\mathbf{k} |
    \Sigma(\varepsilon^{\mathrm{DFT}}_{n\mathbf{k}}) - V_{\mathrm{xc}}
    | n\mathbf{k} \rangle,
\end{equation}
where $\varepsilon^{\mathrm{DFT}}_{n\mathbf{k}}$ is the DFT Kohn-Sham eigenvalue, $V_{\mathrm{xc}}$ the DFT exchange-correlation potential, $\Sigma$ the GW self-energy, and $Z_{n\mathbf{k}}$ is the renormalization factor.  
In 2D semiconductors, the weak dielectric screening results in large self-energy corrections that strongly modify the DFT band structure.
In particular for monolayer SnS$_2$, the GW QP shifts display a non-rigid $\mathbf{k}$-dependence near the M point, where the valence-band saddle plays a crucial role in the nature of the optical excitations.
The bound excitons are then obtained by solving the effective two-particle Hamiltonian \cite{sang+19jpcm,stri88rnc,rich+16book} 
\begin{equation}
    H^{\mathrm{BSE}} \ket{\lambda}
    =
    E_\lambda \ket{\lambda},\,\,
    \ket{\lambda}=\sum_{v,c,\mathbf{k}}\,A^\lambda_{vc\mathbf{k}}\ket{vc\mathbf{k}}
\label{eq:BSE-eigenvalues}    
\end{equation}
where
\begin{equation}
    H^{\mathrm{BSE}}
    =
    (E^{\mathrm{QP}}_{c\mathbf{k}} - E^{\mathrm{QP}}_{v\mathbf{k}})
    \delta_{vv'}\delta_{cc'}\delta_{\mathbf{k}\mathbf{k}'}
    +
    K_{vc\mathbf{k},v'c'\mathbf{k}'}.
\end{equation}

Here, the kernel $K$ contains the electron-hole interaction at the level of the Hartree plus statically screened exchange approximation, capturing, respectively, local field effects and the screened, nonlocal electron-hole attraction responsible for bound states.
The eigenvectors $\ket{\lambda}$ describe the excitonic wavefunctions and they can be expanded in the product basis of valence and conduction states ($vc\mathbf{k}$). The anisotropic distribution of the basis-change coefficients $A^\lambda_{vc\mathbf{k}}$ around the inequivalent M points is central to the characterization of the saddle excitons in monolayer SnS$_2$.
The optical activity of an exciton $\lambda$ for a light polarization direction $\hat{\mathbf{q}}$ is governed by the  excitonic dipole matrix elements
\begin{equation}
    D^\lambda (\hat{\mathbf{q}}) = \sum_{\mathbf{k}} D^\lambda_{\mathbf{k}}(\hat{\mathbf{q}})
    =
    \sum_{v,c,\mathbf{k}}
    A^\lambda_{vc\mathbf{k}}
    \langle v\mathbf{k} \,|\, \hat{\mathbf{q}}\!\cdot\!\mathbf{r} \,|\,
    c\mathbf{k} \rangle.
\label{eq:exc-dipole}
\end{equation}

The $\mathbf{k}$-resolved squared modulus $|D^\lambda_{\mathbf{k}}(\hat{\mathbf{q}})|^2$ reveals how different regions of the BZ contribute to the optical excitation $\lambda$.  
By varying the polarization direction $\hat{\mathbf{q}}$, we show that linearly polarized light selectively excites specific M saddles, enabling three linearly independent excitonic states per exciton $\lambda$. 
In 3D bulk crystals, the optical absorption spectrum is proportional to the imaginary part of the macroscopic dielectric function. However, this quantity is ill-defined for thin films with a nonperiodic direction, and one has to use the 2D polarizability \cite{cuda+11prb,cuda+25prb}:
\begin{equation}
    \alpha_{\mathrm{2D}}(\omega)=
    -\frac{2}{A}\sum_{\lambda} 
     \left|D^\lambda (\hat{\mathbf{q}})\right|
^2 \frac{1}{\omega -E_\lambda +\mathrm{i}\eta},
    \label{eq:alpha2D}
\end{equation}
where $A$ is the unit cell area and $\eta$ an infinitesimally positive value.
In this case, the absorption spectrum is given by $\mathrm{Im}\{\alpha_{\mathrm{2D}}(\omega)\}$.
At the level of the dipole approximation, the $D^\lambda$ matrix elements can also be used to estimate the ``intrinsic'' radiative lifetime $\tau_\lambda$ of bright exciton states (ignoring thermal effects and nonradiative scattering pathways)\cite{palu+15nl,chen+19prb}:
\begin{equation}
1/\tau_\lambda =
\frac{2\pi e^2 E_\lambda}{\hbar^2 c^2 A}
\left|
D^\lambda (\hat{\mathbf{q}})
\right|^2 
    \label{eq:lifetimes}
\end{equation}

\section{Computational details}
\label{compdetails}

\begin{figure}[htb]
    \centering
    \includegraphics[width=1.0\linewidth]{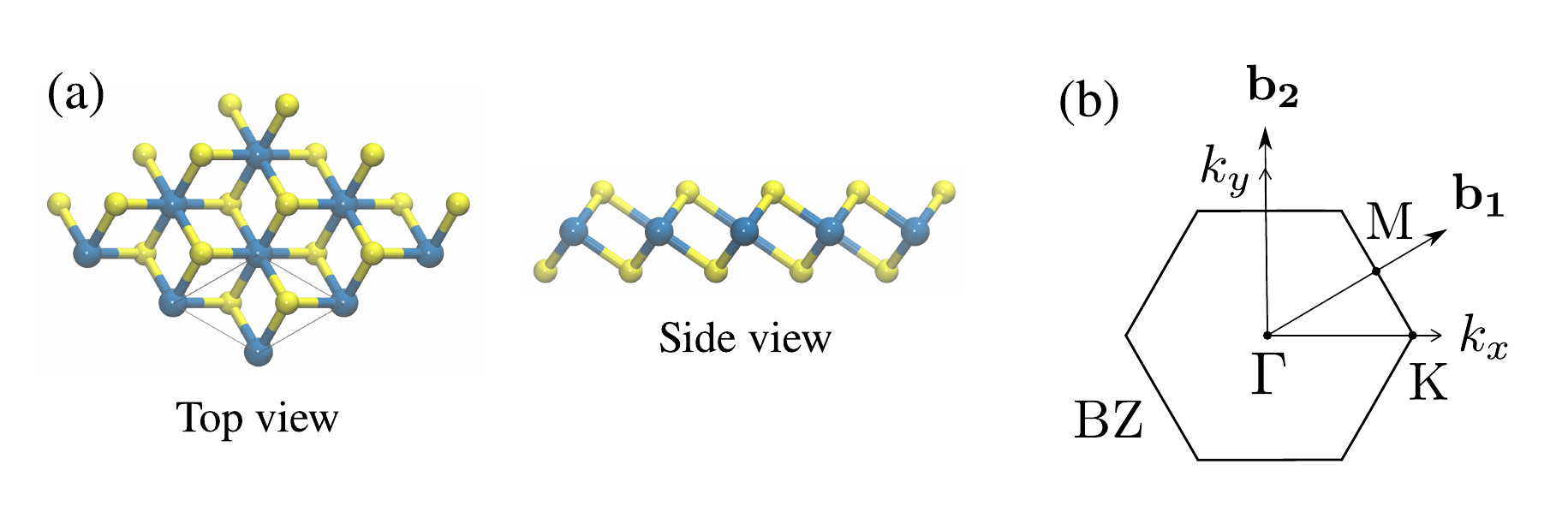}
    \caption{\label{SnS2-geometry} (a) Top and side views of the T-phase monolayer SnS$_2$. 
    The gray line delineates the unit cell. Atomic species (color): Sn (blue) and S (yellow). 
    (b) BZ of the hexagonal Bravais lattice and high-symmetry points used in the band-structure plots.}
\end{figure}

DFT calculations were performed using the Quantum Espresso (QE) package \cite{gian+09jpcm,gian+17jpcm} 
within the PBE approximation \cite{perd+96prl} for the exchange-correlation functional.
The T-phase monolayer SnS$_2$ was described within a periodic-slab model with a vacuum region of 15 {\AA} along the out-of-plane direction.
The interlayer Coulomb interaction was also truncated to remove spurious interlayer screening \cite{sohi+17prb}.
We adopted scalar-relativistic Optimized-Norm-Conserving Vanderbilt Pseudopotentials (ONCVPSP, version 2.1.1) \cite{hama13prb} from the SG15 library \cite{schl-gygi15cpc}, thus neglecting spin-orbit coupling, as it does not affect the band edges of T-phase monolayer SnS$_2$ \cite{baca+16prb}. 
A cutoff energy of 100 Ry (400 Ry) was used for the planewave expansion of the wavefunctions (electronic density).
The GS electronic density was converged using an 8$\times$8$\times$1 Monkhorst-Pack k-point mesh and a minimal SCF convergence threshold of $10^{-8}$~Ry.
The same threshold was adopted at the NSCF iterative diagonalization of the DFT Hamiltonian to generate the input GS band structure for the GW method. 
The starting atomic structure was taken from Ref.~\cite{baca+16prb} and fully relaxed using the default convergence criteria of the QE package. 
In Figure \ref{SnS2-geometry}(a), we show the top and side views of the relaxed monolayer. 
The structure displays a hexagonal Bravais lattice with a lattice parameter of 3.70 {\AA} and a vertical separation of 2.96 {\AA} between the sulfur sublayers.
In Figure \ref{SnS2-geometry}(b), we show the corresponding first BZ and the high-symmetry points used in the band-structure plots.

The QP band structure and, consecutively, the bound excitons were computed using the state-of-the-art many-body approach: the single-shot GW method on top of the DFT GS, followed by the solution of the BSE as implemented in the Yambo code \cite{sang+19jpcm}.
According to the existing ab initio studies of SnS$_2$ layered systems discussed before, DFT-PBE has been demonstrated to provide a good starting-point band structure and wavefunctions to obtain their excited-state properties using many-body perturbation theory, which justifies the single-shot GW approach.
We used a consistent truncation method of the interlayer Coulomb interaction also at this level.
The low-energy bound excitons (below the indirect band gap, E$_{\text{ind}}$) were converged using a reciprocal-space sampling of 48$\times$48$\times$1, and the 5 highest-valence and 2 lowest-conduction bands were included in the BSE Hamiltonian, Eq.~\eqref{eq:BSE-eigenvalues}. 
The convergence of the QP band gap with the k-mesh sampling is generally less demanding \cite{guan+24prb}, and was achieved with 12$\times$12$\times$1.
We employed 340 bands (13 occupied) for the dielectric matrix and the GW self-energy, and a planewave cutoff of 12 Ry for the dielectric matrix. 
In the BSE calculations, the static dielectric matrix requires a much smaller cutoff, and 4 Ry was sufficient to converge the absorption spectrum.
The absorption spectrum was obtained for the in-plane light polarization (1, 1, 0) and using a Lorentzian broadening of 0.01 eV. 
We point out that since monolayer SnS$_2$ has a hexagonal symmetry, the linear optical response is isotropic with respect to the in-plane polarization direction.

\section{Results and Discussion}

\begin{figure}[htb]
    \centering
    \includegraphics[width=1.0\columnwidth]{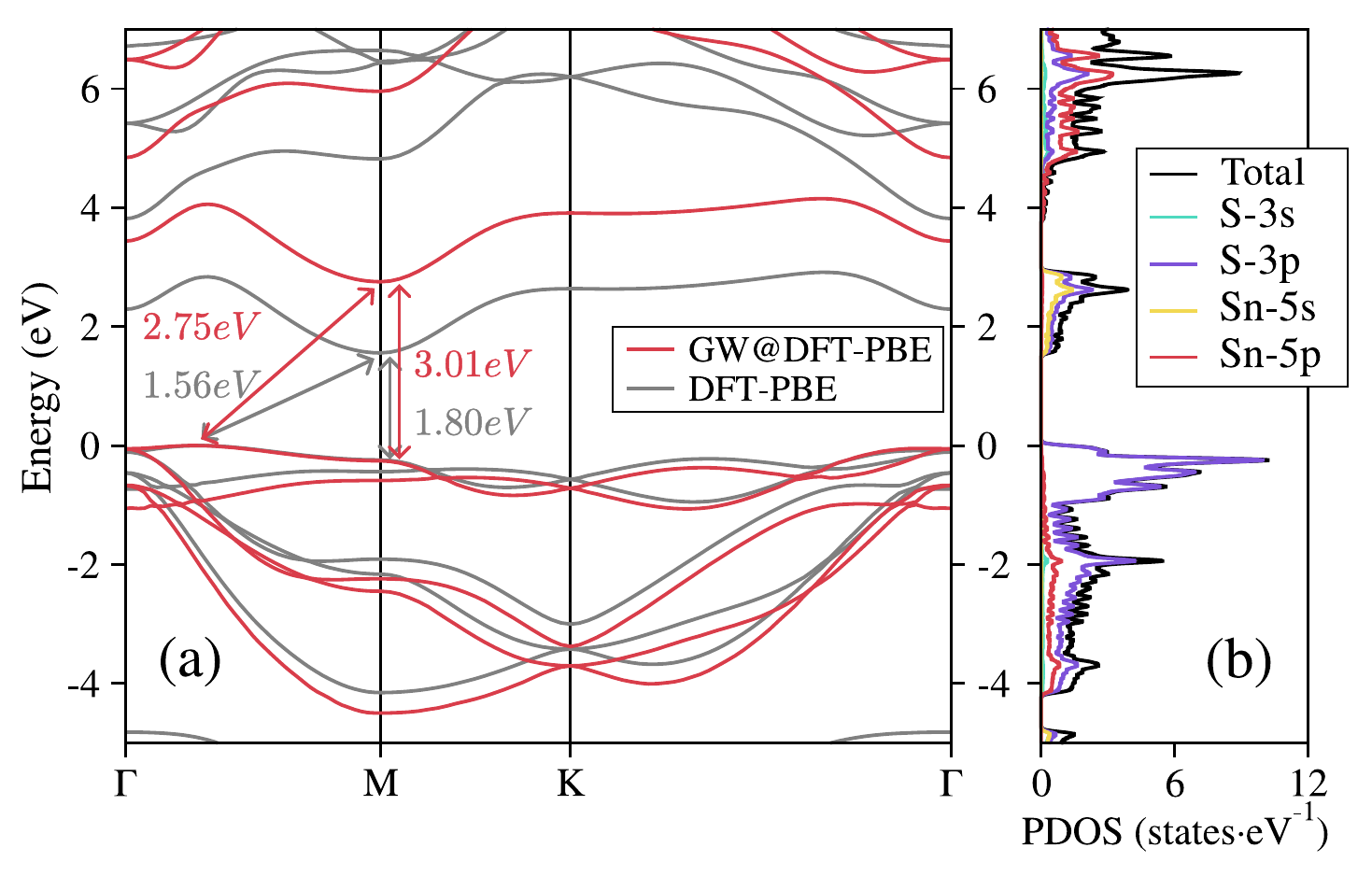}
    \caption{\label{SnS2-bands-PDOS} Band structure of T-phase monolayer SnS$_2$ (a) and DFT-PBE PDOS (b). 
    The arrows in (b) indicate the indirect and direct band gaps.}
\end{figure}

The electronic structure results are shown in Figure \ref{SnS2-bands-PDOS}, where panel (a) displays the DFT and GW band structure plots for monolayer SnS$_2$ along the $\Gamma$MK$\Gamma$ high-symmetry path.
As previously reported in the literature \cite{zhua+13prb,baca+16prb,gonz-oley16prb}, it is an indirect band gap material with valence band maximum placed along $\Gamma$M and conduction band minimum at the M point. 
The band structure plot shows E$_{\text{dir}}$ at M, which is 0.24 eV (0.26 eV) higher in energy than E$_{\text{ind}}$ in the DFT (GW) results, in good agreement with the previous works.
As expected, the GW correction significantly increases E$_{\text{dir}}^{\text{DFT}}$ by 1.21 eV, since DFT tends to severely overestimate screening in 2D materials due to its semilocal-static description of the electron-electron interaction, which is particularly inadequate in low-dimensional systems.
The GW correction also distorts the valence band dispersion due to the momentum dependence of the electronic self-energy, which cannot be properly accounted for by a simple scissor operator. 
Accordingly, we adopted the GW band structure as a starting point for the BSE calculations. 

In Figure \ref{SnS2-bands-PDOS}(b), we also present the DFT projected density of states (PDOS). 
The S-3p atomic orbitals dominate the highest valence band (HVB), with maximum amplitude at 0.25 eV below the top, which is the van Hove singularity associated with the inflection point at M.
Interestingly, the lowest conduction band (LCB), which is formed by a hybridization of S-3p and Sn-5s orbitals, is isolated from the other CBs and exhibits a small bandwidth of around 1.2 eV. 
As we show below, these two bands make the dominant contribution to the bound excitons \cite{fils+16dt,yu+18prb}.

\begin{figure}[htb]
    \centering
    \includegraphics[width=1.0\columnwidth]{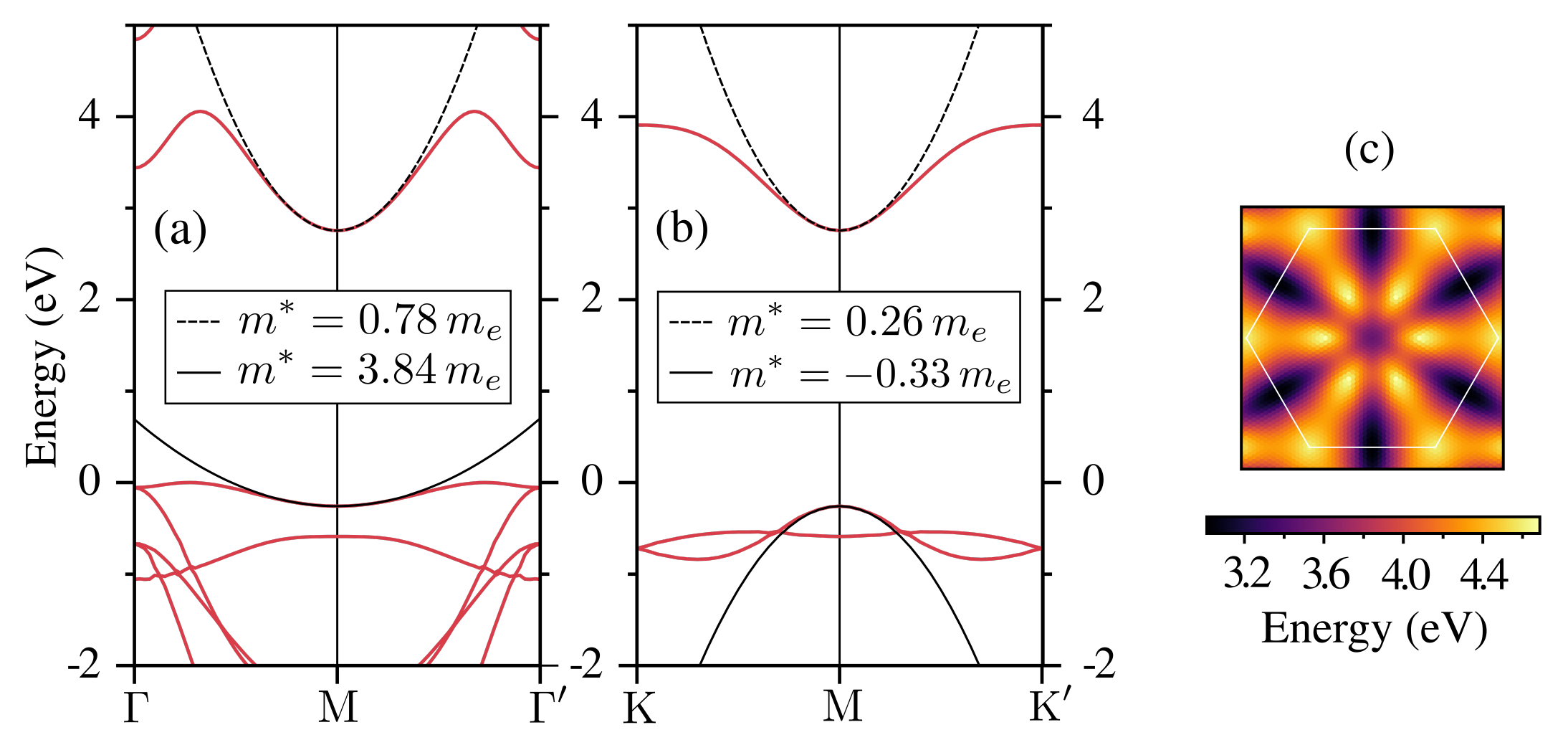}
    \caption{\label{SnS2-QP-bands-saddlepoint} GW@DFT-PBE band structure of T-phase monolayer SnS$_2$ along $\Gamma$M$\Gamma^\prime$ (a) and KMK$^\prime$ (b). 
    The black solid (dashed) lines are the fitted parabolas used to estimate the hole (electron) effective masses $m^*$, which are given in terms of the rest mass of the electron $m_e$. 
    Panel (c) reports the energetic landscape of the GW@DFT-PBE direct band gaps across the BZ, which shows the geometry of the lowest energy regions around $\Gamma$ and M.}
\end{figure}

Focusing now on the band dispersion around E$_{\text{dir}}$, Figure \ref{SnS2-QP-bands-saddlepoint} shows a zoom in the gap region of the GW@DFT-PBE band structure along $\Gamma$M$\Gamma^\prime$ (a) and KMK$^\prime$ (b); the insets report the numerical values of the corresponding electron and hole effective masses at M.
It is evident that the charge carriers are significantly heavier along M$\Gamma$ than along MK, indicating relevant direction-dependent conductivity around the M point.
In the M$\Gamma$ direction, both valence and conduction effective masses have the same sign, with the valence one being almost four times larger ($3.84 \ m_e$ and $0.78 \ m_e$, respectively).
Along MK, the valence effective mass is more than ten times lower than along M$\Gamma$, and comparable to the conduction value ($-0.33 \ m_e$ and $0.26 \ m_e$, respectively).
The change in concavity of the HVB along M$\Gamma$ and MK creates a saddle point at M, a common feature in hexagonal lattices.
However, the fact that it corresponds to the lowest-energy direct electronic transition is quite less common, and it leads to the formation of bound excitons with special symmetry properties, as we discuss below.

There is also a strong connection between the HVB and LCB landscape and the provenance of the single-particle contributions to the bound excitations, which must originate from the lowest energy regions.  
In Figure \ref{SnS2-QP-bands-saddlepoint}(c), we present the landscape of the GW direct band gap across the BZ, showing that the lowest energy region forms an oval shape due to the strong anisotropy around M. 
As we show next, this oval-shaped region is a fingerprint in the bound excitations. 
The secondary minimum around $\Gamma$ also gives a contribution to the higher energy states.

\begin{figure}[htb]
    \centering
    \includegraphics[width=1.0\columnwidth]{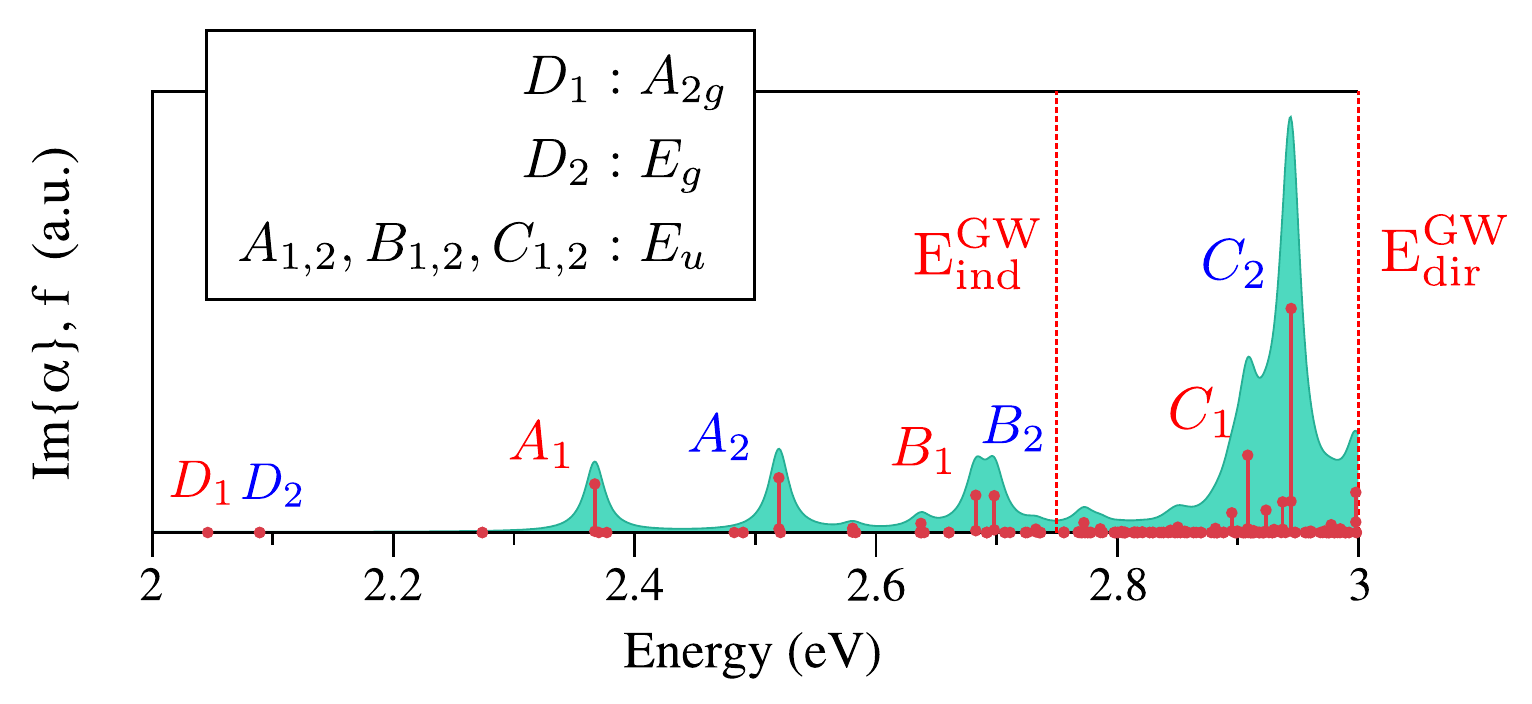}
    \caption{\label{SnS2-BSE} BSE optical absorption spectrum of monolayer SnS$_2$ for in-plane (x, y) electric-field polarization: red dots and bars indicate the optical excitation energies and associated oscillator strengths, the light-green curve is the imaginary part of the 2D polarizability with a Lorentzian broadening of 0.01 eV, and the dotted lines indicate the direct and indirect GW band gaps. 
    The inset shows the symmetries of the excitons in terms of irreducible representations of the $D_{3d}$ point group.}
\end{figure}

Moving now to the optical excitations, Figure \ref{SnS2-BSE} shows the BSE absorption spectrum (light-green curve), Eq.~\eqref{eq:alpha2D}, along with the associated oscillator strengths $|D^\lambda|^2$ (red dots and bars) for all bound excitons.
In Figure \ref{IPCE} of Appendix A, we also show a comparison of the BSE spectrum with the experimental IPCE of a PEC cathode based on monolayer SnS$_2$ \cite{sun+12ac}, which reveals a similar trend between the absorption intensity per wavelength and the amount of generated current.
Our results are not intended to exactly reproduce the experimental data, as many important effects -- such as the dielectric environment, temperature broadening, and charge-separation and collection efficiencies -- are neglected.
Nonetheless, the IPCE onset reported at 540 nm (2.3 eV) is in very good agreement with the onset of the BSE spectrum when adopting a Lorentzian broadening of 0.1 eV to simulate the experimental resolution. 
From this comparison, it also emerges that the features observed in the experimental IPCE spectrum in the 420-540 nm range can be associated with the peaks assigned to bright excitons.

As we notice in Figure \ref{SnS2-BSE}, monolayer SnS$_2$ exhibits a rich structure of excitations in the visible range. 
The first two excitons are dark ($D_1$ and $D_2$), with large binding energies of 0.96 and 0.91 eV, respectively.
Such large binding energies have been predicted before by Gonzalez and Oleynik \cite{gonz-oley16prb} using an effective-mass model for 2D systems, appropriately modified to account for the electronic anisotropy around M.
Remarkably, ab initio calculations now confirm that model predictions, but also elucidate that such lowest excitons are actually dark.
Next, three pronounced absorption peaks emerge below E$_{\text{ind}}^{\text{GW}}$, due to the presence of four bright excitons ($A_1$, $A_2$, $B_1$, and $B_2$) -- at variance with prediction based on simplified hydrogen-like models \cite{zhua+13prb,gonz-oley16prb}, this sequence of peaks presents similar intensity and a somewhat constant energy separation of about 0.2 eV.
Finally, we notice that the spectrum is dominated by a very broad peak between the indirect and direct band gaps, mainly due to two bright excitons ($C_1$ and $C_2$).

In the inset of Figure \ref{SnS2-BSE}, we also present the irreducible representations of the $D_{3d}$ point group according to which the excitonic wavefunctions transform.
The symmetries of excitonic states in a realistic crystal are not simply given by the composition of the symmetries of a fundamental electron-hole transition, like in isolated systems, but must be determined by computing the transformation properties of the full exciton wavefunction, since an exciton is given by a weighted linear combination of all possible electron-hole transitions. This calculation was done using the methodology described in Ref.~\cite{murali+25arxiv}, and implemented in the Yambopy package \cite{yambopy} (note that the symmetry analysis of the single-particle electron-hole transitions is also reported in Appendix B). 
We see that all bright excitons transform according to the $E_u$ representation, like the in-plane polarization vector $\hat{\mathbf{q}}$. Otherwise, by symmetry selection rules, they must be dark.

\begin{figure*}[htb]
    \centering
    \includegraphics[width=0.8\linewidth]{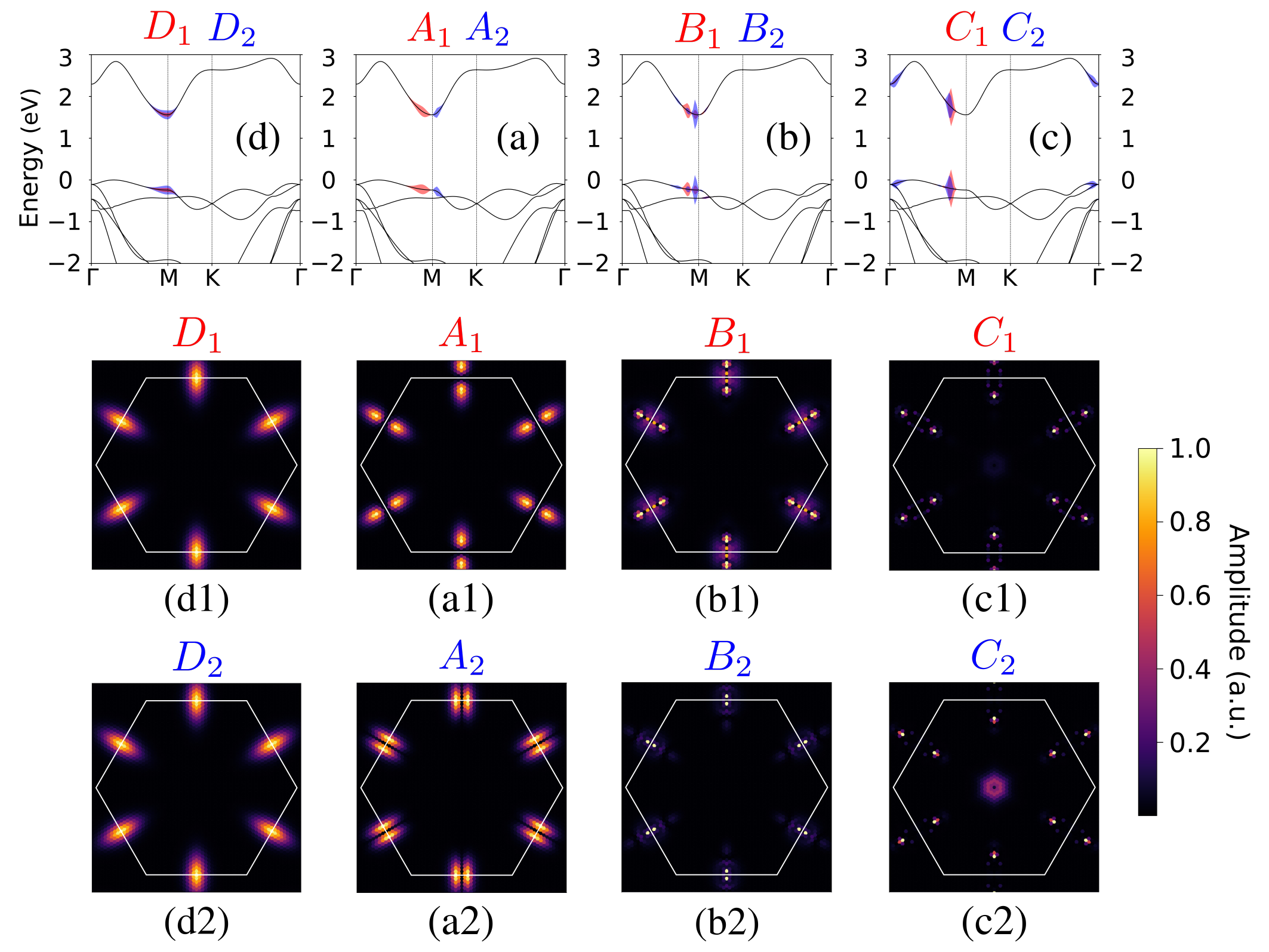}
    \caption{\label{SnS2-exc-on-kspace} Single-particle contributions to the excitons in Figure \ref{SnS2-BSE}: projected on the DFT band dispersion plot ($\sum_{v/c}|A^\lambda_{vc\mathbf{k}}|^2$) along $\Gamma$MK$\Gamma$ (a)-(d), and on the BZ ($\sum_{v,c}|A^\lambda_{vc\mathbf{k}}|^2$) (a1)-(d1) and (a2)-(d2).
    The color maps were independently renormalized by their respective maximum values.}
\end{figure*}

To better qualify the nature of the excitons discussed above, we analyze in Figure \ref{SnS2-exc-on-kspace} the excitonic coefficients $|A^\lambda_{\mathbf{k}cv}|^2$ projected on the DFT band dispersion plot (a)-(d) and on the BZ (a1)-(d1) and (a2)-(d2). 
These coefficients represent the contribution of each single-particle transition $(\mathbf{k}cv)$ to an exciton state $\lambda$.
As seen in Figures \ref{SnS2-exc-on-kspace}(a) to (d), although a reliable convergence of the absorption spectrum below $E_{\text{dir}}$ requires 5 occupied ($v$) and 2 unoccupied ($c$) bands, the dominant electron-hole contributions come from the HVB and LCB.
Nonetheless, it is important to emphasize that the excitons should still be considered as collective excitations, both in k- and band-space, with the energy determined by the BSE Hamiltonian, and cannot be interpreted as a simple single-particle transition.
Interestingly, we notice that the single-particle transitions at exactly $\mathbf{k}=\text{M}$ and $\mathbf{k}=\Gamma$ do not contribute to the bright excitons, as more clearly seen in the BZ projections, whereas they do for the dark states.
Indeed, this is a lattice symmetry requirement.
The $E_u$ ($D_{3d}$) representation of the bright excitons reduces to $A_u \oplus B_u$ ($C_{2h}$) at M.
On the other hand, the HVB and LCB representations at M are $B_g$ and $A_g$, respectively, as shown in the band symmetry analysis in Figure \ref{bands-symmetry} of Appendix B.
Accordingly, they can only contribute to the dark excitons $D_1$ and $D_2$, with representations $A_{2g}$ ($D_{3d}$) $\rightarrow$ $B_g$ ($C_{2h}$) and $E_g$ ($D_{3d}$) $\rightarrow$ $A_g \oplus B_g$ ($C_{2h}$).
At $\Gamma$, the HVB and LCB representations are $E_{g}$ and $A_{1g}$, the latter being the trivial representation.
Consequently, these single-particle transitions could only contribute to dark excitons of the $E_{g}$ representation.

As seen in Figures \ref{SnS2-exc-on-kspace}(d1) and (d2), the dark excitons $D_1$ and $D_2$ exhibit similar single-particle contributions around M, with an oval shape, due to the band dispersion anisotropy in this region, as discussed above.
This same fingerprint is observed in the single-particle contributions to the bright excitons in (a1)-(c1) and (a2)-(c2), however, the contributions are split by nodal planes either along KMK$^\prime$ (A$_1$, B$_1$, B$_2$, C$_1$, C$_2$), or along $\Gamma\text{M}\Gamma^\prime$ (A$_2$).
Furthermore, the excitons $C_1$ and $C_2$ also incorporate additional contributions coming from a second set of transitions around $\Gamma$, as seen in \ref{SnS2-exc-on-kspace}-(c1) and (c2), although the contribution to $C_1$ is very weak.
As shown in Figure \ref{SnS2-BSE-dipoles} of Appendix C, the transition dipoles around $\Gamma$ are noticeably stronger than those around M: this explains the larger oscillator strengths of the $C$-peaks, and in particular of $C_2$, since in this case the contribution from around $\Gamma$ is considerable.

\begin{figure*}[htb]
    \centering
    \includegraphics[width=1.0\linewidth]{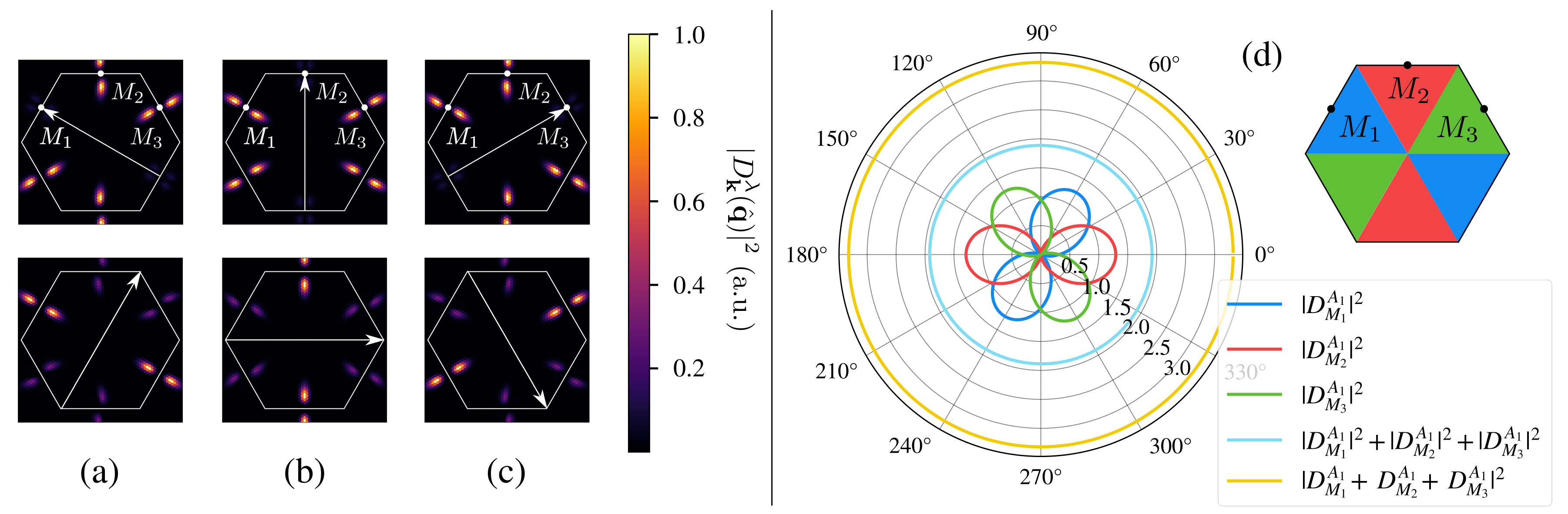}
    \caption{\label{SnS2-A1-dipoles} In (a)-(c), norm-squared of the k-resolved BSE dipoles as a function of light-polarization direction for the first bright exciton $A_1$ ($|D^{A_1}_{\mathbf{k}}(\hat{\mathbf{q}})|^2$). 
    The white arrows indicate the in-plane polarization direction, and the white dots, the high-symmetry inequivalent M points: $M_1$, $M_2$, and $M_3$. 
    On the top figures, the polarization is along $\Gamma$-$M_1$ (a), $\Gamma$-$M_2$ (b), and $\Gamma$-$M_3$ (c); and, on the bottom, along the respective perpendicular directions. The color maps were independently renormalized by their respective maximum values. 
    In (d), a polar plot of $|D^{A_1} (\hat{\mathbf{q}})|^2$ in Eq.~\eqref{eq:exc-dipole} (in units of $10^{-2}\times\text{bohr}^2$), partially integrated over the colored regions shown on the top right, and over the full BZ.}
\end{figure*}

Finally, in Fig.~\ref{SnS2-A1-dipoles}(a)-(c) we show the $k$-resolved polarization-dependent excitonic dipole matrix elements $|D^\lambda_{\mathbf{k}}(\hat{\mathbf{q}})|^2$, Eq.~\eqref{eq:exc-dipole}, for the first bright exciton $A_1$, by adopting different light polarization directions $\hat{\mathbf{q}}$ (white arrows).
The labels $M_1$, $M_2$, and $M_3$ designate the three inequivalent ({\it i.e.} not connected by a reciprocal lattice vector) but symmetry-related ({\it i.e.}  connected by 3-fold rotation) M points.
Focusing on Figure \ref{SnS2-A1-dipoles}(a), as we notice by comparison of the top and bottom panels, the transition dipoles around $M_1$ are orthogonal to the $\Gamma$-$M_1$ direction.
On the other hand, the dipoles around $M_2$ and $M_3$ exhibit non-vanishing components along this direction, which provides the selection of the single-particle excitations around two of the three inequivalent M points.
As expected by the lattice symmetry, the same behavior is seen in Figures \ref{SnS2-A1-dipoles}(b) and (c) regarding the dipoles around $M_2$ and $M_3$, which are perpendicular to the $\Gamma$-$M_2$ and $\Gamma$-$M_3$ directions, respectively.
In Figure \ref{SnS2-A1-dipoles}(d), we also show a polar plot of $|D^{A_1} (\hat{\mathbf{q}})|^2$ in Eq.~\eqref{eq:exc-dipole}, partially integrated over the colored regions shown on the top right, and over the full BZ.
This confirms that, although the linear absorption spectrum, which depends on the fully integrated dipole (Eq.~\eqref{eq:alpha2D}), is ultimately independent of the light-polarization direction due to the crystal hexagonal symmetry (yellow line), linearly polarized light can independently select the single-particle transitions around two of the three inequivalent M points, consequently, giving rise to three linearly independent excitonic states. 
This property has already been theoretically predicted previously in Ref.~\cite{shar+25acsnano}, for light-saddle coupling in graphene. 
However, unlike graphene~\cite{park+17nano,park+21npj2dma}, the saddle point in the band structure of monolayer SnS$_2$, which appears in the valence band at the minimum direct gap, leads to the formation of bound excitons around the M point with well-defined energies and intense optical response.

The same light-polarization selection rule is observed for the remaining bright excitons indicated in Figure \ref{SnS2-BSE}, also with single-particle contributions from along $\Gamma$M, as shown in Figure \ref{SnS2-BSE-dipoles}. 
The only exception is the $A_2$ exciton, for which, on the contrary, the exciton dipoles around $M_i$ ($i=1,2,3$) are directed along $\Gamma M_i$.
As seen by comparison of Figure \ref{SnS2-exc-on-kspace}(a1) and (a2), $A_2$ stems from single-particle contributions along MK, which, although exhibiting electronic densities and symmetry similar to those along $\Gamma$M, have a different distribution, which leads to a different effective-dipole direction.

To gain a more quantitative understanding of the optical activity of the bright excitons in SnS$_2$, we compute their intrinsic radiative lifetimes according to Eq.~\eqref{eq:lifetimes}. 
The resulting values are $\tau_{A_1}=3.9$ ps, $\tau_{A_2}=32$ ps, $\tau_{B_1}=4.3$ ps, $\tau_{B_2}=240$ ps, $\tau_{C_1}=1.9$ ps, and $\tau_{C_2}=13$ ps. 
Notably, these lifetimes are at least one order of magnitude longer than those reported for prototypical transition metal dichalcogenide monolayers~\cite{palu+15nl}, which are widely investigated for valleytronic applications~\cite{scha+16natrm}. 
This comparison highlights a qualitatively different dynamical regime: unlike in graphene, where optical excitation predominantly generates free charge carriers~\cite{shar+25acsnano}, the formation of bound excitons in monolayer SnS$_2$ is associated with relatively long radiative lifetimes. 
Such extended lifetimes provide a temporal window that may enable controlled exciton manipulation prior to recombination, with potential implications for exciton-based optoelectronic functionalities.

\section{Conclusions}

We have presented a detailed analysis of bound excitons in monolayer SnS$_2$ described in the framework of the GW-BSE many-body perturbation theory.
While the M saddle is a common feature in 2D hexagonal lattices, in monolayer SnS$_2$ it coincides with the lowest energy direct single-particle transition, which leads to strongly anisotropic bound excitons with special symmetry properties.
Our BSE spectrum reveals a richer structure of bound excitons than previously reported, likely due to the removal of the spurious interlayer Coulomb interaction in this work.
We find two strongly bound dark excitons, with binding energies of about 0.9 eV, and six strong bright excitons in the visible range, from 2.3 (green) to 3.0 eV (violet).
The first four bright excitons, which are below E$_{\text{ind}}$, originate from near the M saddles in the valence band and give rise to three absorption peaks with similar intensities and equally spaced in energy, in contrast with the behaviour expected from Wannier-Mott excitons (predicting, in the simplest case, $1/n^2$ energy separations and $1/n^3$ peak intensity scaling).
Regarding the two bright excitons lying between E$_{\text{ind}}$ and E$_{\text{dir}}$, they include additional single-particle transitions from near $\Gamma$, with much larger transition dipoles, and accordingly, give rise to a very intense absorption peak about 2.9 eV.
Due to symmetry selection rules, the single-particle transitions at M and $\Gamma$ do not contribute to the bright excitons, which exhibit nodal planes either along KMK$^{\prime}$ or along $\Gamma$M$\Gamma^{\prime}$.
Also, the strong band dispersion anisotropy of the HVB and LCB near M leads to an anisotropic dispersion of the excitonic wavefunctions.
Finally, we have demonstrated that linearly polarized light breaks the $C_{3}$ rotational symmetry between the three inequivalent M points and leads to three linearly independent excitonic states, which can be exploited for state representation in valleytronics.
We hope this work provides a description of the excited-state properties of SnS$_2$ which may prove useful in light of new optoelectronic applications of this material.

\section*{Acknowledgements}

V.A.B., M.G., and A.R. acknowledge the project ECOSISTER ``Ecosystem for Sustainable Transition in Emilia-Romagna'' (ECS$\_$00000033, CUP E93C22001100001) funded by the European Union – NextGenerationEU through the Italian National Recovery and Resilience Plan (PNRR), Mission 4, Component 2, Investment 1.5, also through the V.A.B. PhD scholarship. 
F.P. acknowledges funding by ICSC - Centro Nazionale di Ricerca in High Performance Computing, Big Data and Quantum Computing – funded by the European Union through the Italian Ministry of University and Research under PNRR M4C2I1.4 (Grant No. CN00000013). 
M.G.\ and A.R.\ acknowledge financial support from the EMPEROR project (Grant No.  E93C24001040001), funded by the  European Union—NextGeneration EU (M4C2INV1.3) through the National Quantum Science and Technology Institute (Spoke 5).
A.R. acknowledges the support of the Italian MUR PRIN project “Biodegradable thin film electronics for massively deployable and sustainable Internet of Things applications” (Project n.2022L4YZS4). 
Partial support by MaX - MAterials design at the eXascale, the European Centre of Excellence, co-funded by the European High Performance Computing Joint Undertaking (JU) and participating countries within the HORIZON-EUROHPC-JU-2021-COE-1 program (grant n. 101093374) is also acknowledged. 

\appendix
\section{BSE spectrum vs experimental data}

In this Appendix, we compare the calculated absorption spectrum of monolayer SnS$_2$ with available experimental data. 
The BSE spectrum, obtained at the BSE@GW@DFT-PBE level, is analyzed against the IPCE spectrum measured for a photoelectrochemical cathode based on monolayer SnS$_2$ deposited on ITO-coated glass. 
This comparison allows us to assess the accuracy of the many-body approach in reproducing the excitonic features and overall spectral shape observed experimentally, even though the comparison is between two different observable quantities, as remarked in the main text.

\setcounter{figure}{0} % Reset figure counter to 0
\renewcommand{\thefigure}{A.\arabic{figure}} % Prefix A. to figures
\begin{figure}[htb]
    \centering
    \includegraphics[width=1.0\linewidth]{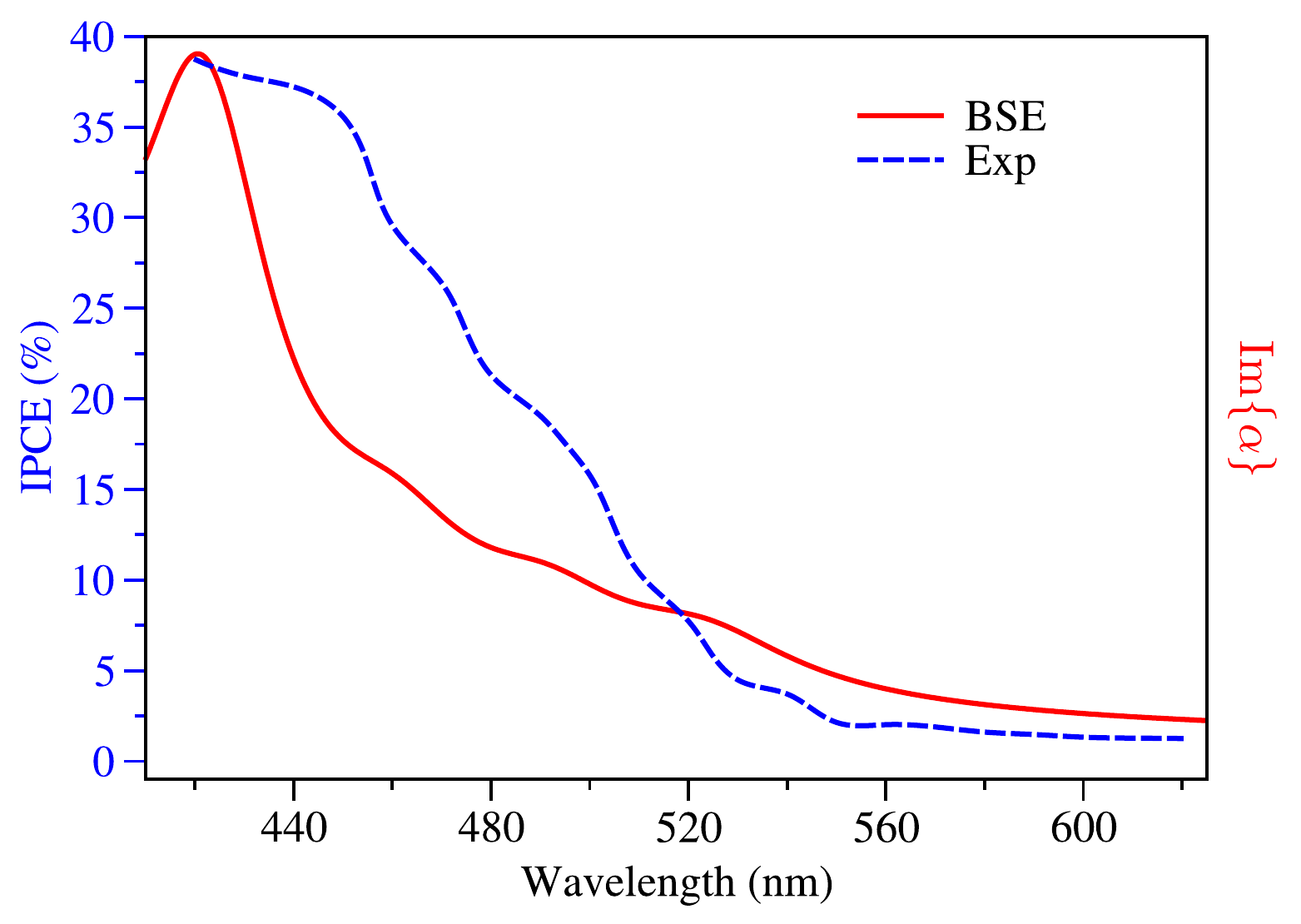}
    \caption{\label{IPCE} IPCE of a photoelectrochemical cathode of monolayer SnS$_2$ deposited on ITO-coated glass \cite{sun+12ac}. The BSE curve was obtained from the BSE@GW@DFT-PBE absorption spectrum of the isolated monolayer SnS$_2$ with a Lorentzian broadening of 0.1 eV, renormalized by the experimental efficiency at 420 nm of 38.7$\%$.}
\end{figure}

\newpage
\onecolumngrid
\section{Band symmetry analysis}

This Appendix presents the symmetry analysis of the electronic bands of monolayer SnS$_2$, based on DFT-PBE calculations. 
In addition to the band dispersion along the $\Gamma$–M–K–$\Gamma$ path, we assign the corresponding point-group representations at the high-symmetry points and along each segment. 
The symmetry character of the HVB and LCB is further illustrated through representative electronic-density isosurfaces, which provide insight into the orbital nature of the band edges.

\setcounter{figure}{0} % Reset figure counter to 0
\renewcommand{\thefigure}{B.\arabic{figure}} % Prefix B. to figures
\begin{figure*}[h!]
    \centering
    \includegraphics[width=1.0\linewidth]{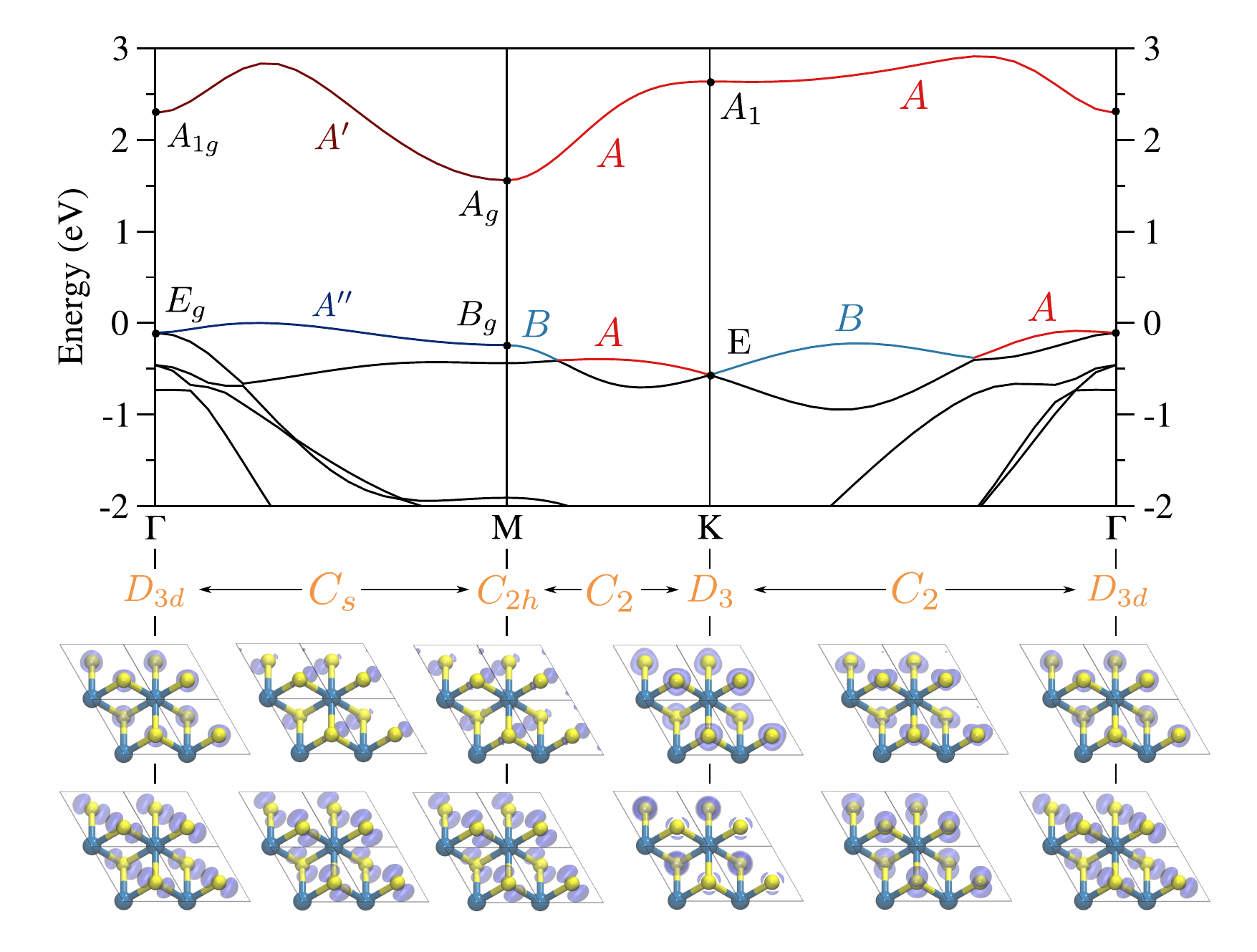}
    \caption{\label{bands-symmetry} DFT-PBE band structure of monolayer SnS$_2$ and symmetry analysis of the HVB and LCB. The point group symmetries at the k points along $\Gamma$MK$\Gamma$ are shown in orange at the bottom, as well as electronic-density isosurfaces of representative orbitals of the HVB (bottom orbitals) and LCB (top orbitals). Isosurface cutoff of 10$^{-2}$ electrons$\cdot$bohr$^{-3}$. The HVB and LCB point-group representations along each path section and at the high-symmetry points are indicated in the band dispersion plot.}
\end{figure*}

\newpage
\setcounter{figure}{0} % Reset figure counter to 0
\renewcommand{\thefigure}{C.\arabic{figure}} % Prefix C. to figures
\section{Exciton dipoles}

In this last Appendix, we present the norm-squared of the k-resolved BSE dipole matrix elements for the bright excitons discussed in Figure 4 of the main text, as a function of the in-plane light-polarization direction. 
The maps highlight the anisotropic distribution of the excitonic dipole strength around the inequivalent M points and provide insight into the symmetry-driven/polarization-dependent selection rules. 
It is also clear that for the $C$-states (especially $C_2$), the contribution around $\Gamma$ is responsible for their larger oscillator strengths. This contribution is not present for the $A$ and $B$-states.

\begin{figure*}[h!]
    \centering
    \includegraphics[width=1.0\linewidth]{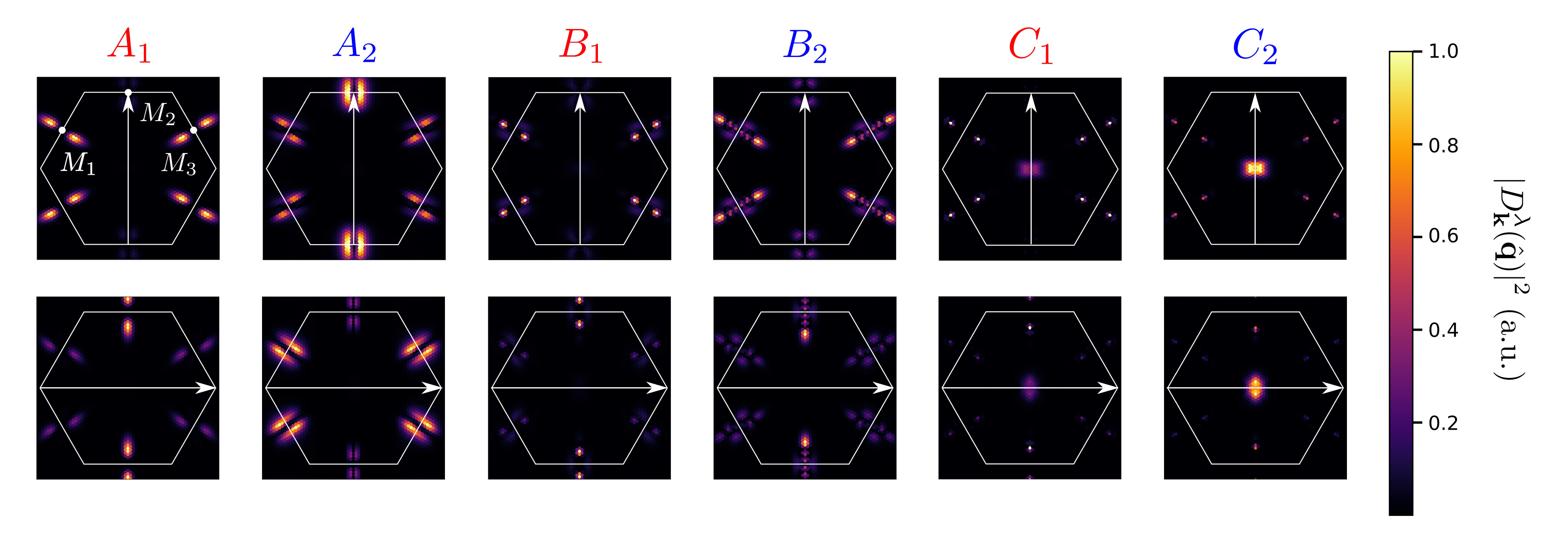}
    \caption{\label{SnS2-BSE-dipoles} Norm-squared of the k-resolved BSE dipoles as a function of light polarization for the bright excitons indicated in Figure 4 of the main text. The white arrows indicate the in-plane light-polarization direction, and the white dots in the top-left graphic, the high-symmetry inequivalent M points: $M_1$, $M_2$, and $M_3$. On the top figures, the polarization is along $\Gamma$-$M_2$, and, on the bottom figures, along the perpendicular direction. The color maps were independently renormalized by their respective maximum values.}
\end{figure*}

\twocolumngrid
\bibliography{bastosetalBIBLIO-NAMES,bastosetalBIBLIO}

\end{document}